\documentclass{aa}
\usepackage{amsmath}
\usepackage[varg]{txfonts}
\usepackage[english]{babel}
\usepackage{graphicx,tikz, wasysym}
\usepackage{caption, setspace, subfigure}
\usepackage{epstopdf,epsfig}
\usepackage{hyperref}
\usepackage{natbib}
\bibpunct{(}{)}{;}{a}{}{,}

\begin{document}

\title{The projected gravitational potential of the galaxy cluster MACS~J1206 derived from galaxy kinematics}
\author{Dennis Stock\inst{1} \and Sven Meyer\inst{1} \and Eleonora Sarli\inst{1} \and Matthias Bartelmann\inst{1} 
 \and Italo Balestra\inst{2} \and Claudio Grillo\inst{3} \and Anton Koekemoer\inst{4} \and Amata Mercurio\inst{5} \and Mario Nonino\inst{2} \and Piero Rosati\inst{6}}
\institute{Universit\"at Heidelberg, Zentrum f\"ur Astronomie, Institut f\"ur Theoretische Astrophysik, Philosophenweg 12, 69120 Heidelberg, Germany
 \and INAF - Osservatorio Astronomico di Trieste, via G. B. Tiepolo 11, I-34131, Trieste, Italy
 \and Dark Cosmology Centre, Niels Bohr Institute, University of Copenhagen, Juliane Maries Vej 30, DK-2100 Copenhagen, Denmark
 \and Space Telescope Science Institute, 3700 San Martin Drive, Baltimore, MD  21218, USA
 \and INAF - Osservatorio Astronomico di Capodimonte, Via Moiariello 16, I-80131, Napoli, Italy
 \and  Dipartimento di Fisica e Scienze della Terra, Università degli Studi di Ferrara, Via Saragat 1, I-44122 Ferrara, Italy
 }

\abstract{We reconstruct the radial profile of the projected gravitational potential of the galaxy cluster MACS~J1206 from 592 spectroscopic measurements of velocities of cluster members. For doing so, we use a method we have developed recently based on the Richardson-Lucy deprojection algorithm and an inversion of the spherically-symmetric Jeans equation. We find that, within the uncertainties, our reconstruction agrees very well with a potential reconstruction from weak and strong gravitational lensing as well as with a potential obtained from X-ray measurements. In addition, our reconstruction is in good agreement with several common analytic profiles of the lensing potential. Varying the anisotropy parameter in the Jeans equation, we find that isotropy parameters which are either small, $\beta\lesssim0.2$, or decrease with radius yield potential profiles which strongly disagree with that obtained from gravitational lensing. We achieve the best agreement between our potential profile and the profile from 
gravitational lensing if the anisotropy parameter rises quite steeply to $\beta\approx0.
6$ within $\approx0.5\,\mathrm{Mpc}$ and stays constant further out.}

\maketitle

\section{Introduction}

Galaxy clusters offer several classes of observables reflecting their overall internal constitution: gravitational lensing effects in their weak and strong variants, X-ray emission, the thermal Sunyaev-Zel'dovich (tSZ) effect and the kinematics of their member galaxies. We neglect radio emission, turbulence and the metal abundance in the intra cluster gas as well as the population statistics of the member galaxies here because these effects are locally driven.

Gravitational lensing measures the gravitational tidal field, projected along the line-of-sight and thus directly probes the projected gravitational potential, more precisely its curvature. The X-ray emission and the thermal Sunyaev-Zel'dovich effect depend on powers of the density and the temperature of the intra\-cluster medium. If equilibrium assumptions hold, hydrostatic and virial equilibrium foremost, the X-ray emission, the thermal Sunyaev-Zel'dovich effect and galaxy kinematics will also be determined by the gravitational potential. Current discussions of the validity of these equilibrium assumptions can be found e.g. in \cite{equilibrium1, equilibrium2, equilibrium3, equilibrium4}.

What is the gravitational potential that agrees best with all cluster observables? This question is relevant for different reasons. First, different observables trace the gravitational potential at different scales. The cluster core can be probed by the stellar kinematics of the brightest cluster galaxy, strong lensing and X-ray emission probe the innermost regions, weak lensing and galaxy kinematics probe large scales, and the tSZ effect falls in between (see \cite{xray1, Merten, bcgdynamics}). Aiming at a reliable reconstruction of cluster density profiles from their cores to their outskirts, combining all observables into a unique potential reconstruction offers the advantage of covering all relevant scales in a single step. Second, comparing cluster potential reconstructions based on lensing on the one hand and based on the rest of the observables on the other hand allows testing the equilibrium assumptions or possible deviations therefrom. Third, lensing and the other observables do not necessarily see 
the same gravitational potential. While lensing is sensitive to the sum of the Bardeen potentials, the other observables probe the 
spatial potential only. In general relativity, the two Bardeen potentials agree in case of negligible anisotropic stress. Differences in potential reconstructions based on lensing compared to other observables may also hint at deviations from relativity \citep[see also][in this context]{2014ApJ...783L..11S, Barreira}.

Aiming at the gravitational potential has the major and important advantages that it is a locally measureable quantity which is directly related to the observables listed above \citep[see also][]{2009A&A...494..461A, 2012A&A...538A..98A}.

We have recently developed methods for reconstructing the projected gravitational potential of galaxy clusters from their X-ray emission, their tSZ effect and the kinematics of their member galaxies \citep{galkin, xray, 2013arXiv1304.6522M}. They operate similarly, but with important differences in detail: An observable is deprojected by means of the Richardson-Lucy algorithm, requiring symmetry assumptions. The deprojected quantities are related to the three-dimensional gravitational potential by relations derived from justifiable equilibrium assumptions. The gravitational potential can then be projected along the line-of-sight. These methods complement our techniques for joint cluster reconstruction from weak and strong gravitational lensing (\citealt{1996ApJ...464L.115B, 2006A&A...458..349C, 2009A&A...500..681M}; see also \citealt{2012ApJ...757...22C, 2011MNRAS.417..333M, Julian_meshfree} for examples).

In this paper, we reconstruct the projected gravitational potential of the galaxy cluster MACS~J1206.2$-$0847 based on galaxy kinematics, applying the technique developed in \cite{galkin}. This work is structured as follows: Sect.~2 briefly reviews the reconstruction method. In Sect.~3, we describe the selection and preparation of the data. Sect.~4 presents the results and compares the projected gravitational potential to that obtained from gravitational lensing. We summarise in Sect.~5 and discuss other methods in Sect.~6.

\section{Reconstruction method}

The basic assumptions of our reconstruction method are the following: We treat the galaxy cluster as a spherically symmetric gas cloud of collisionless, pointlike test particles of mass $m$, i.e.~galaxies, moving in the gravitational potential of their common dark matter halo. The system can then be described by the Jeans equation relating the radial velocity dispersion $\sigma_\mathrm{r}^2$ weighted by the galaxy number density $\rho_\mathrm{gal}$ to the gravitational potential $\phi:=\Phi/m$,
\begin{equation}
  \frac{1}{\rho_\mathrm{gal}}
  \frac{\mathrm{\partial}(\rho_\mathrm{gal}\sigma_r^2)}{\mathrm{\partial} r}+
  2\beta\frac{\sigma_r^2}{r} =
  -\frac{\mathrm{\partial}\phi}{\mathrm{\partial}r}\;,
\label{Jeans}
\end{equation}
where the anisotropy parameter $\beta:=1-\sigma_\theta^2/\sigma_r^2$ quantifies the ratio between the tangential and the radial velocity dispersions.

Expanding on the formal analogy with gas dynamics, we define an effective galaxy pressure $P:=\rho_\mathrm{gal}\sigma_r^2$. Using this definition and setting $\beta$ to zero, (\ref{Jeans}) would turn into the equation of hydrostatic equilibrium for a gas. In addition, we introduce a polytropic relation between the effective galaxy pressure and the matter density,
\begin{equation}
  P = P_0\left(\frac{\rho}{\rho_0}\right)^{\gamma}\;.
\label{polytropic}
\end{equation}
This was justified for a simulated cluster in \cite{galkin}. In fact, the authors first tested with several density profiles that the polytropic assumption is reasonable. Additionally, we shall see in our reconstruction results that the exact choice of the polytropic index affects the final results only very mildly.

\begin{figure}
\begin{center}
\begin{tikzpicture}
  \node (A) at (0,2.5) {\Large{$\rho_\mathrm{gal}\sigma_\mathrm{los}^2$}};
  \node (B) at (4,2.5) {\Large{$\rho_\mathrm{gal}\sigma_r^2$}};
  \node (C) at (4,0)   {\Large{$\Phi$}};
  \node (D) at (0,0)   {\Large{$\Psi$}};
  \node (E) at (0,1.8) {\small{(Observable)}};
  \node (G) at (4.75,1.25) {\scriptsize{Jeans eq.}};
  \draw[->,thick] (A) to node[above]{\scriptsize{Richardson-Lucy}}node[below]{\scriptsize{deprojection}} (B);
  \draw[->,thick] (B) to (C);
  \draw[->,thick] (C) to node[above]{\scriptsize{Projection}} (D);
\end{tikzpicture}
\caption{Scheme of the reconstruction algorithm. The observable is converted to a three-dimensional quantity by Richardson-Lucy deprojection. This is then turned into the three-dimensional gravitational potential, solving the Jeans equation. The resulting potential is finally projected along the line-of-sight.}
\label{scheme}
\end{center}
\end{figure}
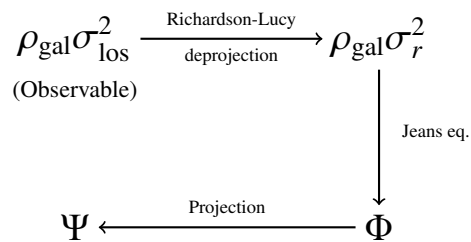

The reconstruction algorithm described in detail in \cite{galkin} proceeds along the following steps (see also Fig.~\ref{scheme}):
\begin{enumerate}
  \item In order to obtain the effective pressure $P$, we first have to deproject the actual observable, i.e.~the line-of-sight projected velocity dispersion weighted by the galaxy number density, $\rho_\mathrm{gal}\sigma_\mathrm{los}^2$. This is done via the Richardson-Lucy deconvolution or deprojection, see \cite{Lucy74,Lucy1994}.
  \item Furthermore, we make use of the polytropic relation (\ref{polytropic}) to rewrite the Jeans equation (\ref{Jeans}) in terms of the effective pressure. This leads to a Volterra integral equation of the second kind for the gravitational potential $\Phi$. After fixing the shape of the anisotropy profile, we can solve this equation in a quickly converging iteration process.
  \item Finally, we project $\Phi$ along the line-of-sight to find the projected gravitational potential $\Psi$.
\end{enumerate}

This entire algorithm depends on three parameters and a function being carefully adapted during the reconstruction. Noise suppression within the Richardson-Lucy deprojection requires regularisation, controlled by a smoothing scale $L$ and an amplitude $\alpha$. Furthermore, we have to choose the polytropic index $\gamma$ in (\ref{polytropic}), usually being of order unity, as cluster analyses in \cite{galkin} suggest. This also implies that the galaxy fluid can approximately be treated as an isothermal gas. The importance of the polytropic index is investigated in more detail in Sect. 4.1.

As we shall demonstrate later, the dominant parameter is the anisotropy profile $\beta(r)$. However, if kinematic data are used exclusively for cluster reconstruction, there is in principle a degeneracy between the anisotropy parameter and the gravitational potential. We resolve this well-known degeneracy by fixing the 
$\beta$-profile in order to obtain a non-parametric estimate for the gravitational potential. To avoid an arbitrary, unmotivated guess for $\beta(r)$, we will choose it such that the reconstructed potential agrees best with a reconstruction based on gravitational-lensing data. Alternative methods for breaking the anisotropy-mass degeneracy and for recovering the gravitational potential using galaxy kinematics will be dicussed in the last section.

\section{Data selection and preparation}

The observational input data for the relaxed and massive galaxy cluster MACS~J1206 ($M_\mathrm{lens}\approx5\cdot10^{14}\mathrm{M}_\odot h^{-1}$ \cite{mass_macs1206}) investigated here consist of a catalogue of member galaxies \cite{biviano_m1206}, observed in the context of the CLASH project \citep{2012ApJS..199...25P} as part of a large spectroscopic campaign carried out with the Very Large Telescope (VLT) (CLASH-VLT Large Programme; \cite{rosati_clashvlt}). According to \cite{biviano_m1206} and \cite{radialincompleteness_annunz}, the spatial incompleteness of the spectroscopic sample varies with position in the cluster by less than $20\%$. 

The following analysis is based on $592$ member galaxies in total, selected via the method presented in \cite{biviano_m1206}. In order to arrive at the line-of-sight projected velocity dispersion profile, we assign to every cluster member a projected radius from the cluster centre, which we take to be marked by the brightest cluster galaxy. This is achieved by multiplying the angular separation with the angular-diameter distance, calculated with a standard $\Lambda$CDM-cosmology with recent cosmological parameters from \cite{planck}.

\begin{figure}
\includegraphics[width=\columnwidth]{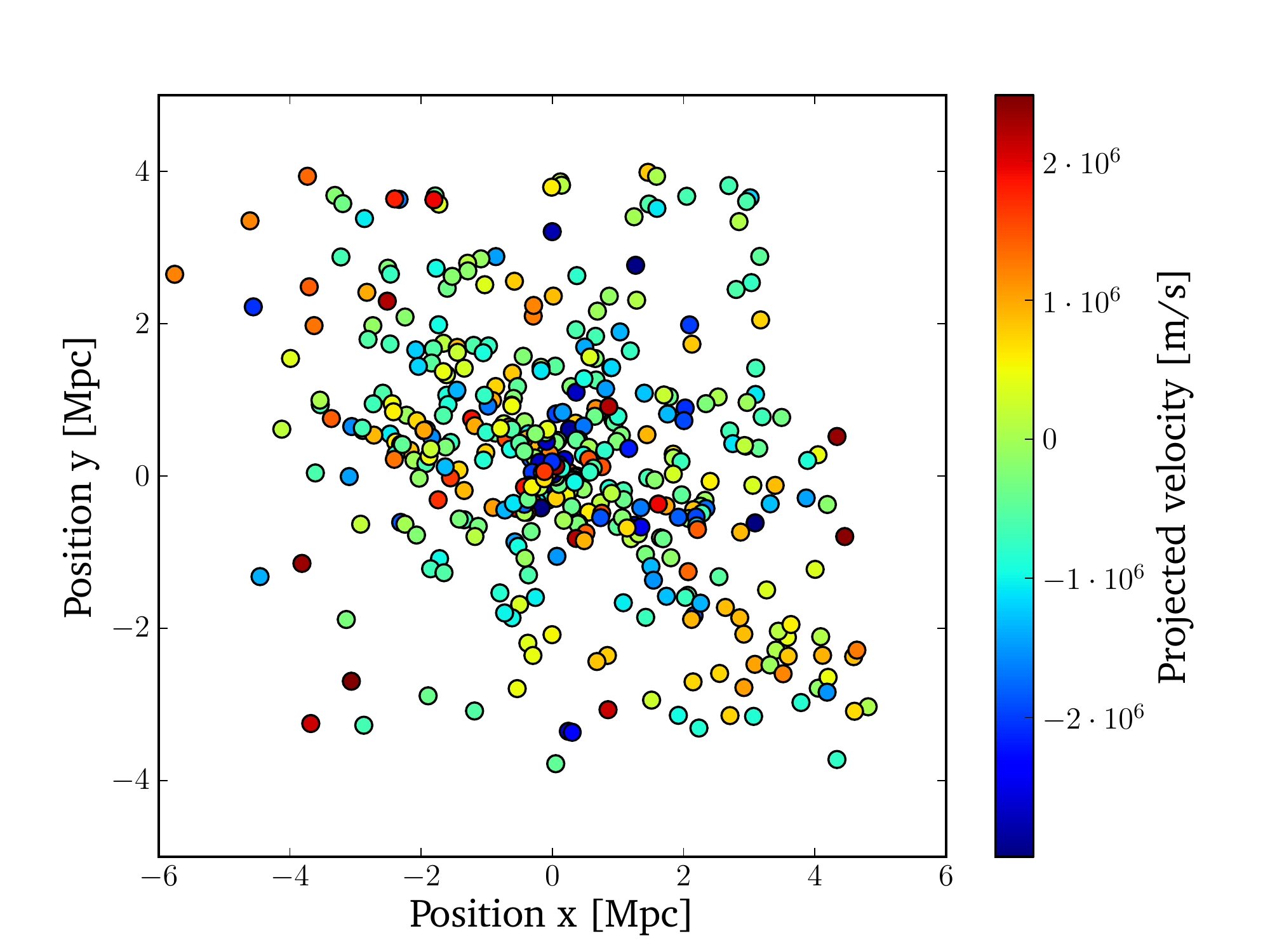}
\caption{Velocities projected along the l.o.s. of cluster members.}
\label{velocities}
\end{figure}

We determine the line-of-sight projected velocities from the measured redshifts as described in \cite{veldis2}. In our case we can use the non-relativistic relation between redshift $z$ and line-of-sight velocity $v_\mathrm{los}$, since all involved velocities (see below) are clearly in the non-relativistic regime. Thus,
\begin{equation}
  v_\mathrm{los}=c\, z\;.
\end{equation}
As we are only interested in the galaxy velocities with respect to the cluster's centre-of-mass, we have to take into account that the observed redshift includes four main contributions: the motion of the observer with respect to a local comoving observer, the motion of the cluster's centre-of-mass with respect to a local comoving observer, the motion of each galaxy with respect to the cluster's centre-of-mass, and finally the Hubble expansion. \cite{veldis2} has shown that neglecting the motion of the cluster's centre-of-mass leads to a multiplication of redshifts,
\begin{equation}
  1+z = (1+z_0)(1+z_\mathrm{cosm})(1+z_\mathrm{G})\;,
\label{redshifts}
\end{equation}
where $z_0$ is the redshift due to the observer's motion with respect to a local comoving observer, $z_\mathrm{cosm}$ represents the Hubble expansion and $z_\mathrm{G}$ the motion of each galaxy with respect to a local comoving observer. To leading order, we can calculate the cluster average of (\ref{redshifts}) and use that the average redshift of all member galaxies with respect to the local comoving frame vanishes because we assume an isotropic distribution of velocities along the l.o.s.,
\begin{equation}
  \langle z_\mathrm{G}\rangle=0\;.
\end{equation}
Furthermore, we can safely assume that the contribution by the Hubble expansion is the same for all cluster members,
\begin{equation}
  \langle z_\mathrm{cosm}\rangle=z_\mathrm{cosm}\;.
\end{equation}
Thus, we end up with
\begin{equation}
  1+z_\mathrm{cosm}=\frac{1+\langle z\rangle}{1+z_0}\;.
\label{zcosm}
\end{equation}
Inserting (\ref{zcosm}) into (\ref{redshifts}), we find the following expression for $v_\mathrm{los}$, see also figure \ref{velocities}:
\begin{equation}
  v_\mathrm{los} = c\, z_\mathrm{G} =c\left(\frac{1+z}{1+\langle z\rangle}-1\right)\;.
\end{equation}

To arrive at the density-weighted, projected velocity dispersion profile $\rho_\mathrm{gal}\sigma_\mathrm{los}^2$, we adapt the bin width to the one chosen for the potential from lensing for better comparability, i.e.~a constant bin width of $0.1\,\mathrm{Mpc}$. Next, we calculate the velocity dispersion within each bin and reconstruct the projected gravitational potential as described above by solving the radial Jeans equation.

Due to the gauge freedom of the lensing potential, one is in particular allowed to add an arbitrary constant. By convention, we scale the potential such that $\Psi(0)=1$ and use the normalisation $\Psi(R_\mathrm{cut})=0$ for a given cut-off radius $R_\mathrm{cut}=3\,\mathrm{Mpc}$. This is even larger than the virial radius of MACS~J1206, being approximately $2\,\mathrm{Mpc}$ \citep{mass_macs1206}.

In order to obtain error bars and to suppress the effect of outliers, we perform a bootstrap analysis, i.e.~for each bootstrap sample, we draw as many times with replacement from the original dataset as it has member galaxies. Then, we reconstruct the gravitational potential for each individual bootstrap sample. This procedure is repeated 300 times. Finally, we calculate the mean of all potentials and their standard deviation. The bootstrap analysis also allows to assess the effects of incompleteness and uncertain membership assignment by varying the sample of galaxies entering into the potential reconstruction.

\section{Results}
\subsection{Effect of reconstruction parameters}

As indicated above, we have to fix three reconstruction parameters in order to perform the reconstruction: the regularisation amplitude $\alpha$, the smoothing scale $L$ and the polytropic index $\gamma$. Additionally, we have to model the anisotropy profile $\beta(r)$.

\begin{figure*}
\begin{center}
\subfigure []{\includegraphics[width=\columnwidth]{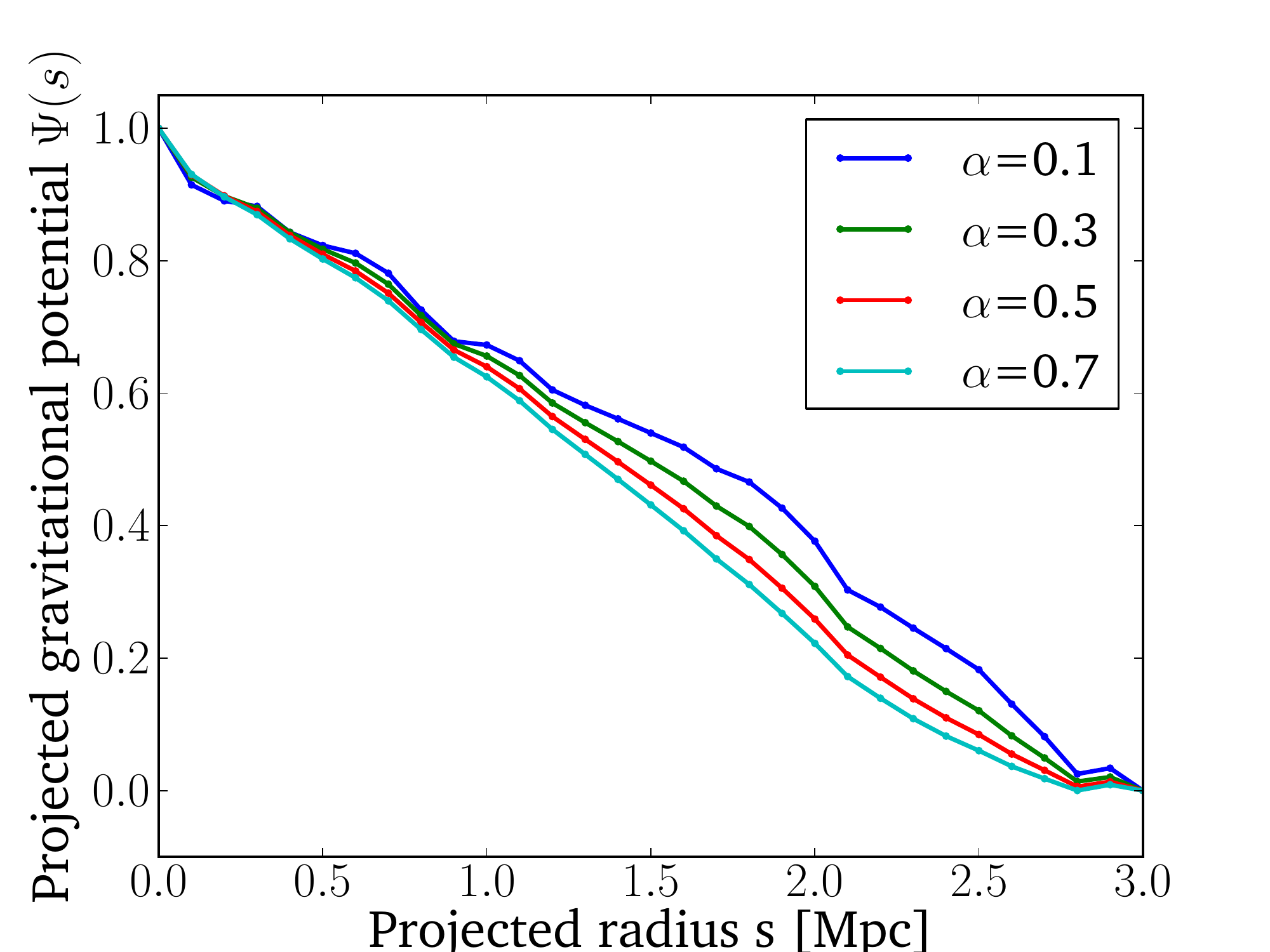}}\hfill
\subfigure []{\includegraphics[width=\columnwidth]{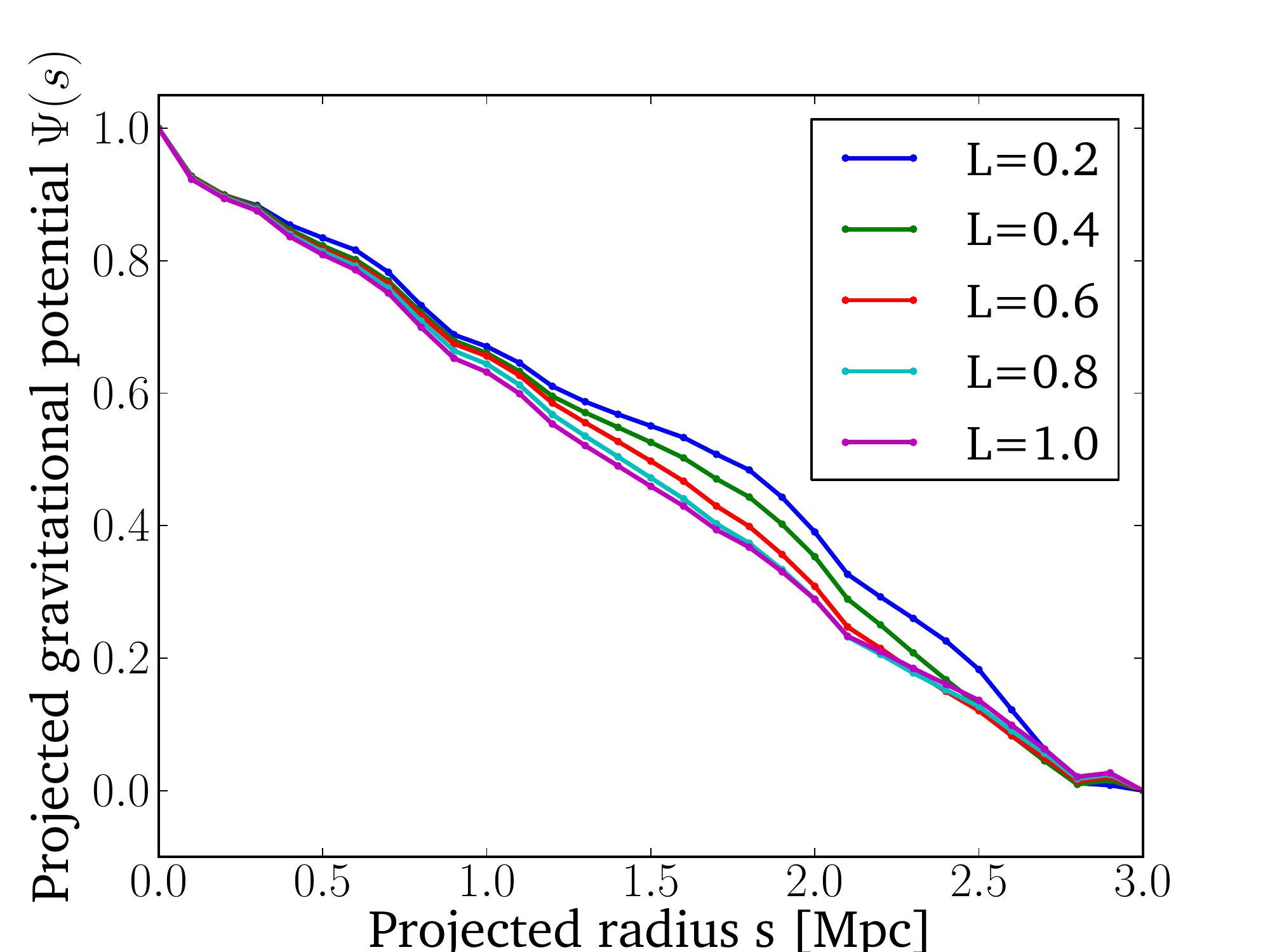}}
\subfigure []{\includegraphics[width=\columnwidth]{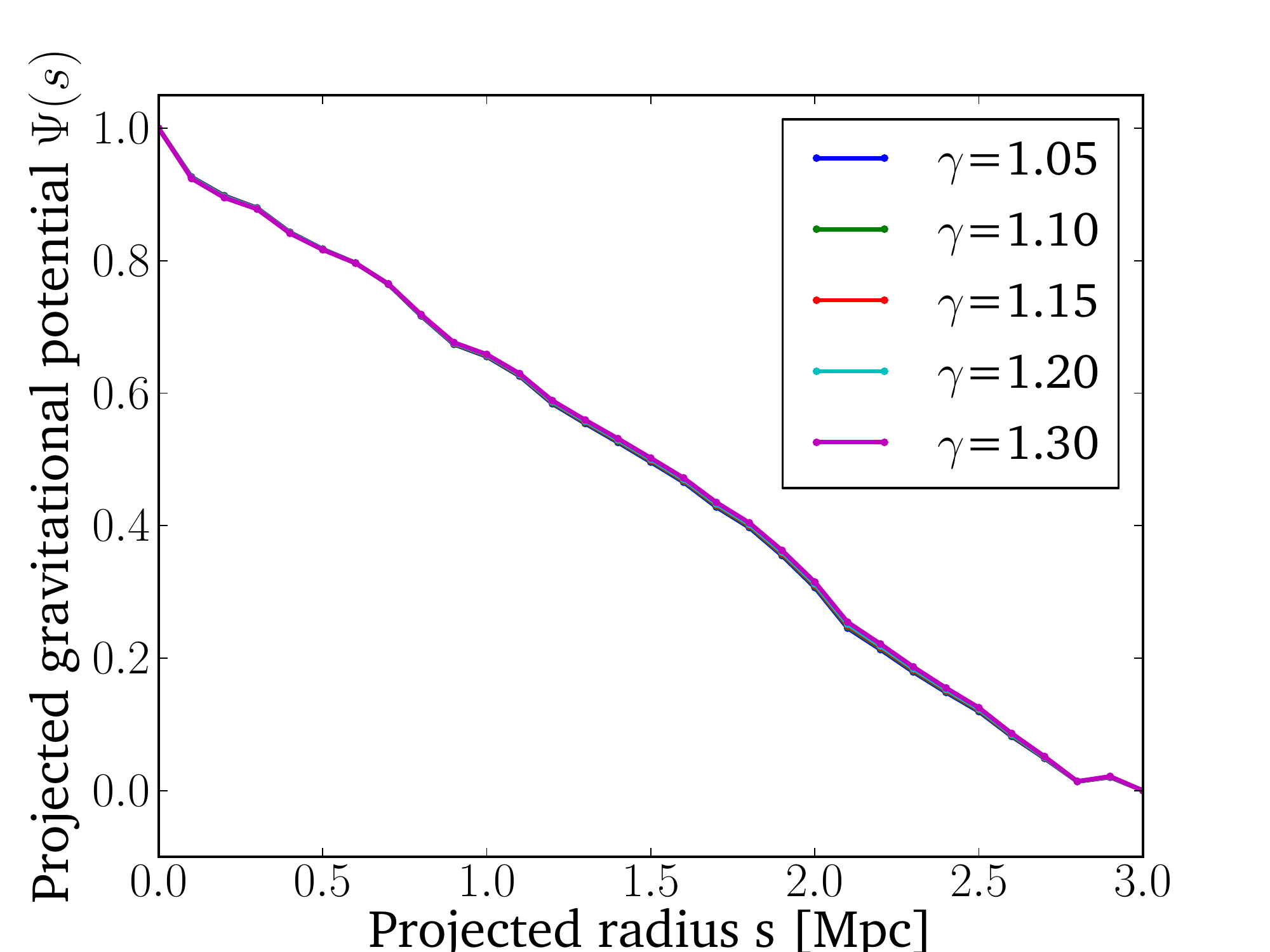}}\hfill
\subfigure []{\includegraphics[width=\columnwidth]{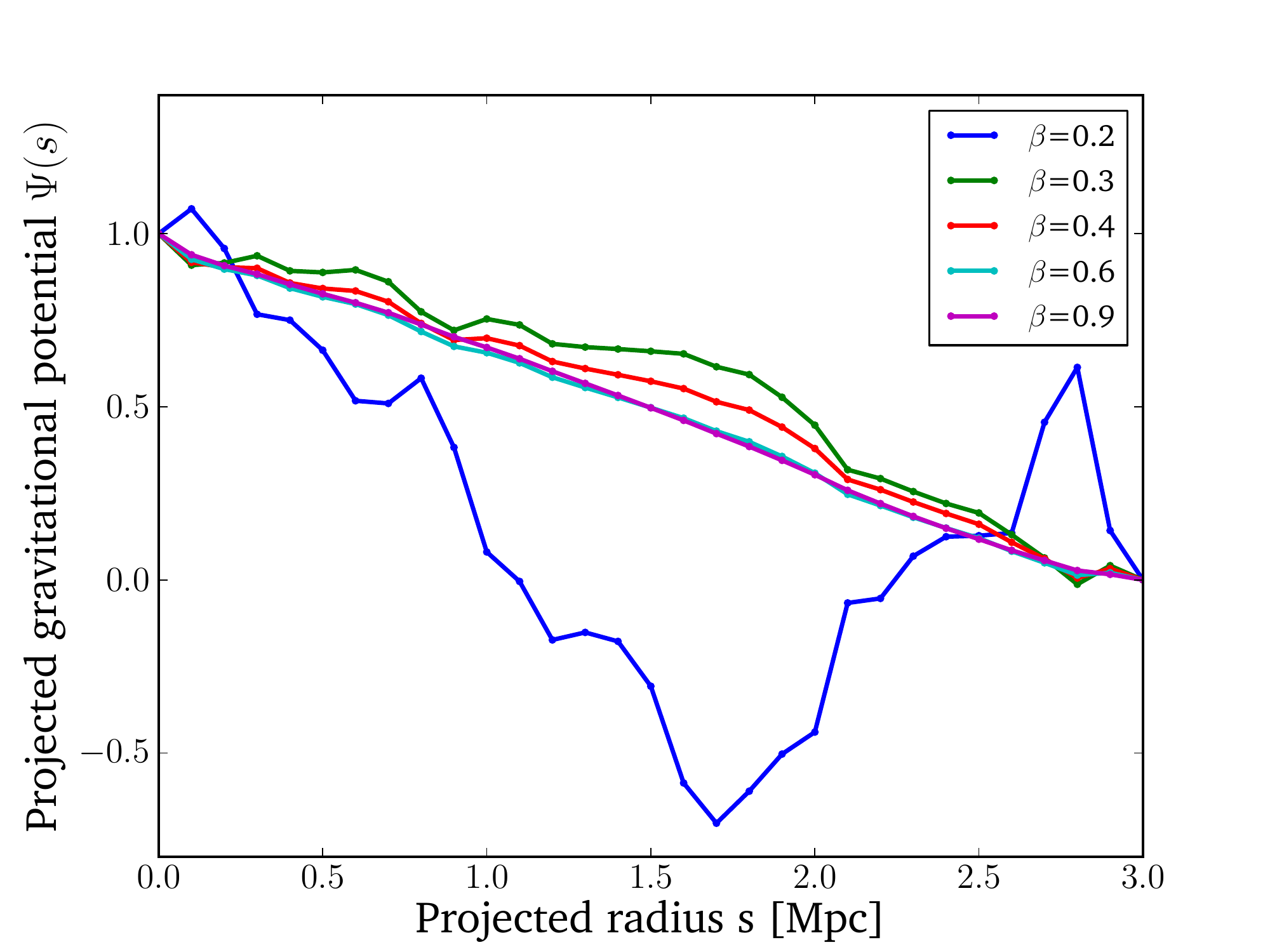}}
\caption{Reconstructed radial profiles of the projected gravitational potential of MACS~J1206, obtained using different parameter sets for the smoothing amplitude $\alpha$ (panel a), smoothing scale $L$ (panel b), the polytropic index $\gamma$ (panel c) and the anisotropy parameter $\beta$ (panel d).}
\label{reconstruction parameters}
\end{center}
\end{figure*}

Figure~\ref{reconstruction parameters} shows how the variation of these reconstruction parameters affects the resulting potential $\Psi$. The amplitude of the anisotropy parameter $\beta$ clearly has the dominant effect. In particular, an isotropic velocity distribution corresponding to $\beta\rightarrow 0$ seems very implausible because of the strongly varying shape of the potential resulting from this assumption.

Although the assumption of a polytropic relation (\ref{polytropic}) may appear quite bold, Fig.~\ref{reconstruction parameters}c shows that the particular choice of the polytropic index does not really matter. Also the effects of the regularisation parameters $\alpha$ and $L$ remain within the uncertainties, being of similar order as shown in Fig.~\ref{lensing potentials}. When they are not varied, the parameters are set as follows: $\alpha=0.3$, $L=0.6$, $\gamma=1.1$ and $\beta=0.6$. The value chosen for the polytropic index $\gamma$ is motivated by the result of \cite{galkin} that $\gamma$ would usually be around unity. The regularisation parameters are chosen such that the agreement with the lensing reconstruction is best.

\subsection{Variable anisotropy profile}
We also systematically investigate the effect of an anisotropy parameter varying with radius. For doing so, we compare radial profiles of gravitational potentials obtained with an anisotropy parameter increasing or decreasing linearly in discrete steps every $0.5\,$Mpc. In case of the decreasing anisotropy-parameter profile, we choose $\beta=0.8$, $0.7$, $0.6$, $0.5$, $0.4$, $0.3$, and reverse for the increasing case.

\begin{figure}
  \includegraphics[width=\hsize]{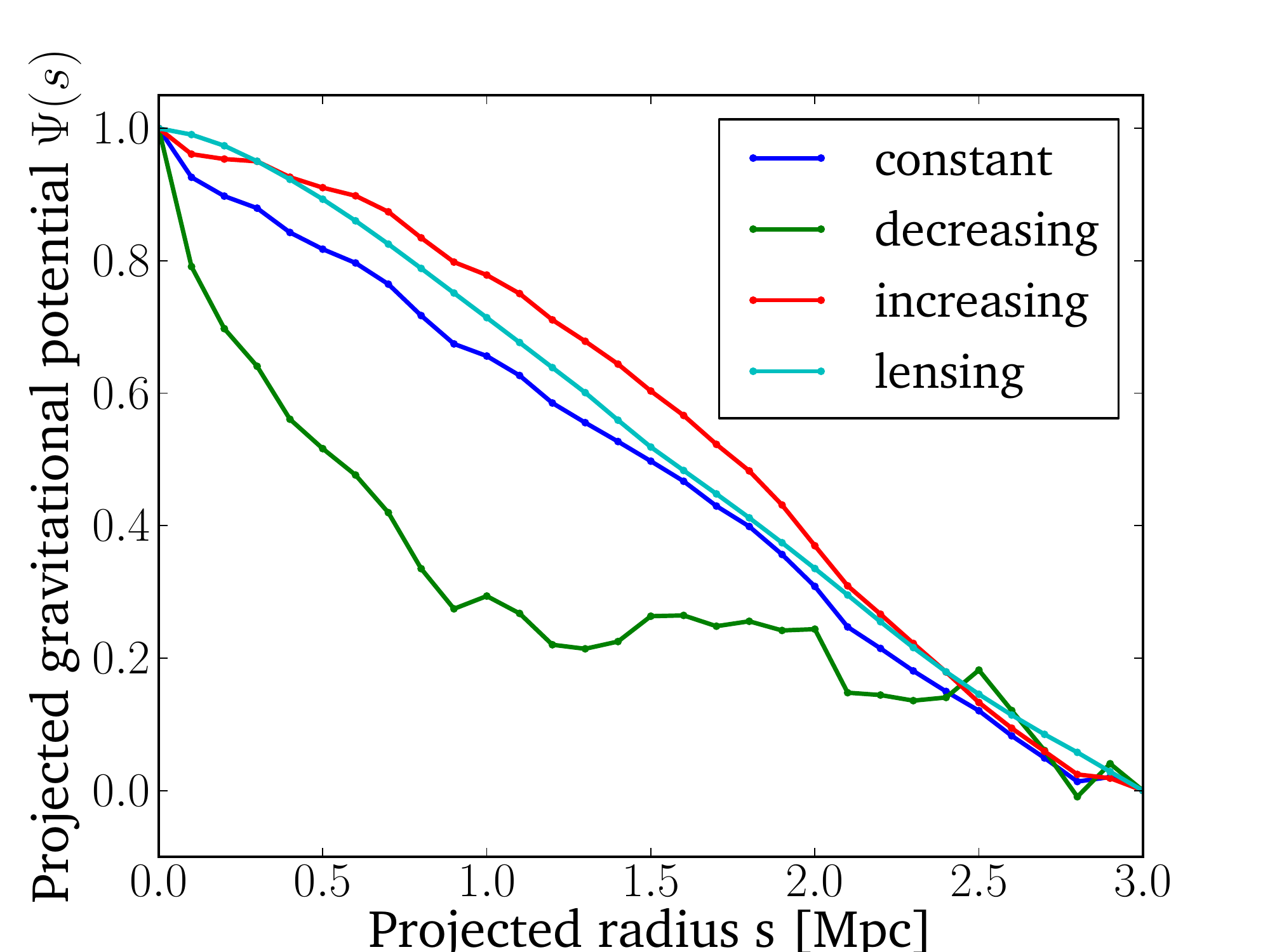}
\caption{Radial profiles of the projected gravitational potential, reconstructed from galaxy kinematics with different profiles of the anisotropy parameter $\beta$. The other reconstruction parameters remain fixed at $\alpha=0.3$, $L=0.6$ and $\gamma=1.1$.}
\label{XXX}
\end{figure}

Figure~\ref{XXX} shows both cases together with the potential reconstruction from weak and strong gravitational lensing and a reconstruction with constant $\beta=0.6$. First, one can observe that the qualitative form of the anisotropy profile affects the curvature of the potential: An increasing $\beta$ profile leads to a potential with negative curvature. In contrast, a decreasing $\beta$-profile leads to a positive curvature and strong variations in the potential at radii beyond $1\,\mathrm{Mpc}$. The reconstruction using a constant $\beta$ leads to an almost straight line for the radial potential profile. 

These results motivate the shape of the anisotropy profile used later in (\ref{beta}). Since the potential from lensing first shows a negative curvature at small radii and a linear evolution later, we choose an anisotropy profile increasing for small radii and turning constant afterwards to reproduce the linear behaviour.

\subsection{Comparison with lensing and X-ray data}
\begin{figure}
\begin{center}
\includegraphics[width=0.5\columnwidth]{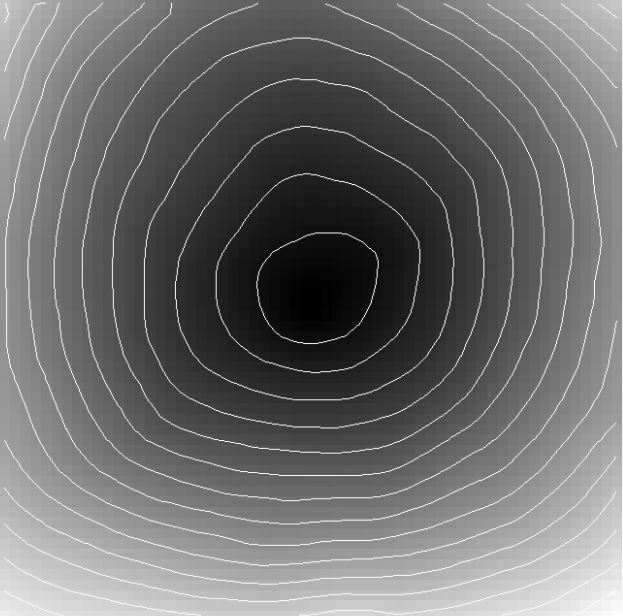}
\caption{Two dimensional map of the lensing potential of MACS1206 based on weak and strong lensing.}
\label{lensing_potential_map}
\end{center}
\end{figure}

\begin{figure}
\includegraphics[width=\hsize]{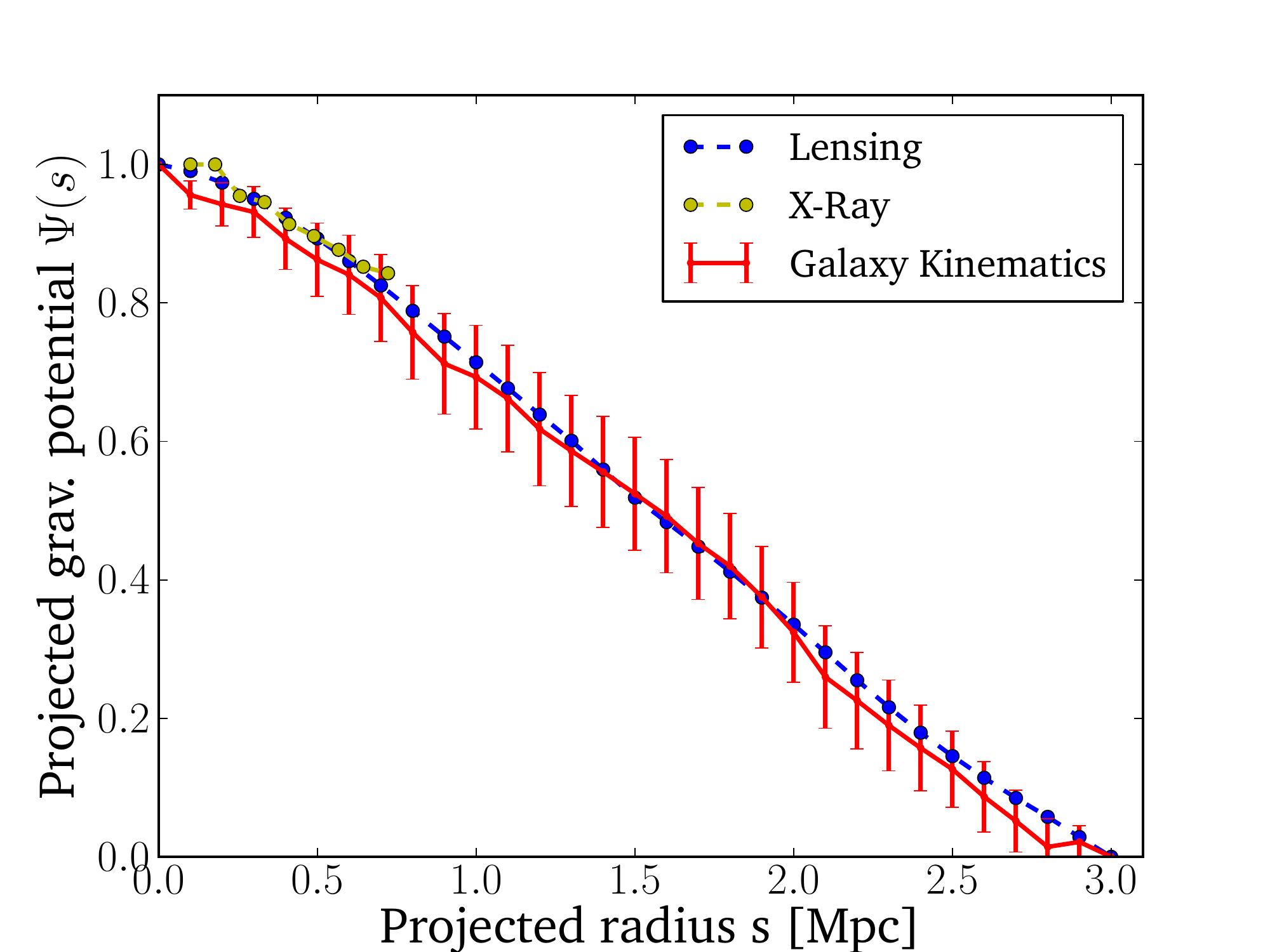}
\caption{Radial profiles of the gravitational potential of MACS~J1206, reconstructed from combined strong and weak gravitational lensing (blue), X-ray emission (yellow) and from galaxy kinematics (red) with the parameters $\alpha=0.3$, $L=0.6$, $\gamma=1.1$ and $\beta$ as in (\ref{beta}).}
\label{lensing potentials}
\end{figure}

As a first test, we compare the projected gravitational potential reconstructed from galaxy velocities with a reconstruction based on weak and strong gravitational lensing data  of \cite{Merten} using constraints from \cite{Zitrin}. After converting the 2D-lensing potential map (see figure \ref{lensing_potential_map}) into a radial profile and using the same normalisation process as above, we can compare the result from gravitational lensing to the potential obtained from cluster kinematics (Fig.~\ref{lensing potentials}).

The reconstruction parameters are chosen as listed above. For the radial profile of the anisotropy parameter $\beta$, we assume the following radially dependent profile, being motivated in Sect. 4.2. This profile turns out to be in qualitative agreement with the one in \cite{biviano_m1206}:
\begin{equation}
  \beta(r)=
  \begin{cases}
    0.1 & r < 0.2 \\
    0.4 & 0.2 \le r <0.5 \\
    0.6 & 0.5 \le r \le 3.0
  \end{cases}\;.
\label{beta}
\end{equation}\\
Additionally, we compare it with a reconstruction based on X-ray emission of the intracluster gas based on \cite{xray_megan} using Chandra data up to $0.72\,$Mpc, which is a typical range of validity using this method for galaxy clusters.\\
The growing difference between the lensing and kinematic profile outside $\sim2\,\mathrm{Mpc}$ can be explained by the limited range of validity of each reconstruction method. The small difference at the innermost radii is due to increasing baryonic effects causing the equilibrium assumption ultimately to break down. The X-ray data agree quite well within the error boundaries.

\subsection{Comparison with common lensing potentials}
\begin{figure}
  \includegraphics[width=\hsize]{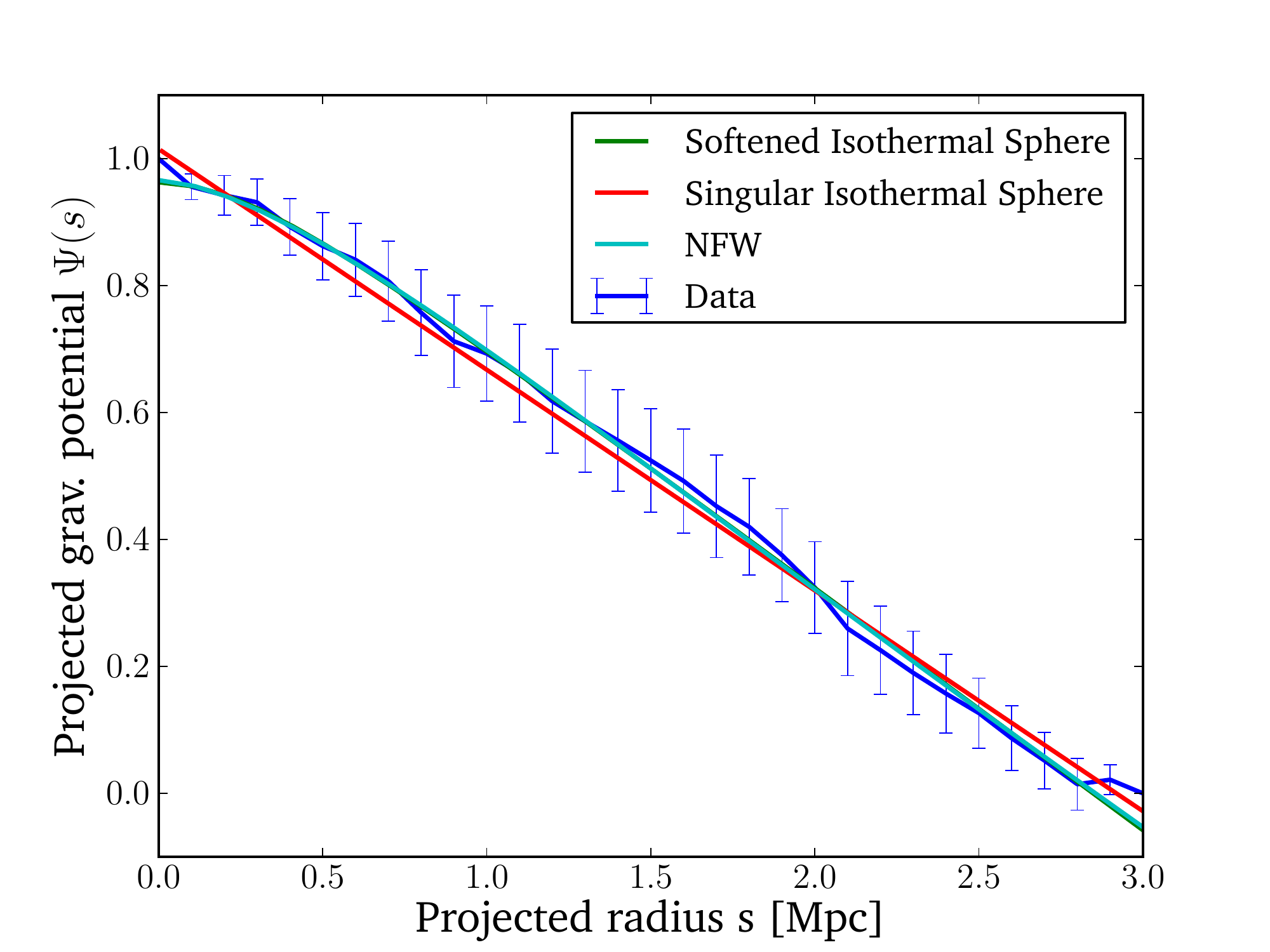}
\caption{Comparison of three analytic profiles of gravitational-lensing potentials with the gravitational-potential profile of MACS~J1206 reconstructed from galaxy kinematics using $\alpha=0.3$, $L=0.6$, $\gamma=1.1$ and $\beta$ as in (\ref{beta}).}
\label{lensing profiles}
\end{figure}

Furthermore, we can compare the radial profile of the projected gravitational potential obtained from galaxy kinematics with radial profiles common in gravitational-lensing studies, such as the singular (SIS) and the non-singular (NIS) isothermal spheres and the NFW profile \citep{NFW, lectgrav}; see Fig.~\ref{lensing profiles}:
\begin{align}
  \Psi_\mathrm{SIS}(s) &= A\cdot s+B \nonumber\\
  \Psi_\mathrm{NIS}(s) &= C\cdot\sqrt{s_\mathrm{c}^2+s^2}+D \nonumber\\
  \Psi_\mathrm{NFW}(s) &= E\cdot h(x)+F
\label{potmodels}
\end{align}
with the function $h(x)$ depending on the radius in units of the scale radius $x:=r/r_\mathrm{s}$ (see \cite{nfwlensing}):
\begin{equation}
h(x) =
\begin{cases}
\displaystyle
\ln^2\frac{x}{2}- \mathrm{arcosh}^2\frac{1}{x} & (x < 1) \\[6pt]
\displaystyle
\ln^2\frac{x}{2}+ \arccos^2 \frac{1}{x}& (x \ge 1)
\end{cases}\;,
\end{equation}
where $s_\mathrm{c}$, $r_\mathrm{s}$ and $A\ldots F$ are free parameters being determined by fitting the models (\ref{potmodels}) to the potential obtained from galaxy kinematics. The parameter values are summarised in Tab.\ref{fit} .
\begin{table*}
\caption{Best fit parameters for different models}
\label{fit}
\centering
\begin{tabular}{c | c c c}
\hline\hline
Model & Parameters \\
\hline
Singular Isothermal Sphere & $A = -0.35\pm 0.01$ & $B = 1.02\pm 0.01$ & \\
Non-singular Isothermal Sphere & $C = -0.39\pm 0.01$ & $s_\mathrm{c} = 0.38\pm 0.06$ & $D = 1.11\pm 0.02$ \\
NFW & $E = -0.84\pm 0.10$ & $r_\mathrm{s} = 1.37\pm 0.15$ & $F = 0.97\pm 0.01$\\
\hline
\end{tabular}
\end{table*}

We can compare the results for the scales $s_\mathrm{c}$ in case of the softened isothermal sphere and $r_\mathrm{s}$ in case of NFW with \cite{mass_macs1206}. They obtain:
\begin{align}
s_\mathrm{c} = 0.000028\pm 0.000006\\
r_\mathrm{s} = 0.345\pm 0.050\; .
\end{align}
The agreement is quantified by a goodness-of-fit parameter $\Omega$, which we define in analogy to the $\chi^2$-function appropriate for uncorrelated measurements. Since our data points are not independent but correlated by the bootstrap method, i.e.~some data points appear multiple times, we cannot interpret $\Omega$ as a $\chi^2$-function. Because we just want to single out the best model, we do not go into a more elaborate analysis for correlated measurements here. For the $i$-th data point $x_i$ with standard deviation $\sigma_i$ and the corresponding model prediction $f(x_i)$, we have
\begin{equation}
  \Omega := \sum_i\left(\frac{f(x_i)-x_i}{\sigma_i}\right)^2\;.
\end{equation}
As one can already see by eye, the softened isothermal sphere and the NFW-model describe the data equally well. More quantitatively, the goodness-of-fit parameters for all lensing model are:
\begin{align}
  \Omega_\mathrm{SIS} &= 6.09 \nonumber\\
  \Omega_\mathrm{NIS} &= 4.00 \nonumber\\
  \Omega_\mathrm{NFW} &= 3.54\;.
\end{align}
So far, all analytic lensing potentials are compatible with the reconstruction from galaxy kinematics, but the error bars are substantial.

\section{Summary}

We have recently developed a method for reconstructing the projected gravitational potential of galaxy clusters from the kinematics of their member galaxies. In the study described in this paper, we have applied this method for the first time to a real galaxy cluster (MACS~J1206). The reconstruction assumes that the three-dimensional gravitational potential of this cluster is approximately spherically symmetric. The potential reconstruction is based on 592 measured galaxy velocities.

Our results can be summarised as follows:
\begin{itemize}
  \item The reconstruction algorithm requires four parameters to be set: the anisotropy parameter $\beta$; the polytropic index $\gamma$ of the effective galaxy pressure; the amplitude $\alpha$ of the regularisation term in the Richardson-Lucy deconvolution and the length scale $L$ of the regularisation. We found that the reconstructed potential is almost insensitive to the ``nuisance parameters'' $\gamma$, $\alpha$ and $L$, while it does depend quite sensitively on the anisotropy parameter $\beta$.
  \item Within the bootstrap error bars, the reconstructed potential profile is indistinguishable from the potential profile obtained from the combination of weak and strong gravitational lensing and from X-ray analysis. Typical analytic lensing profiles, such as the singular and non-singular isothermal spheres and the NFW profile also agree well with our reconstruction from galaxy kinematics.
  \item Small anisotropy parameters $\beta\lesssim0.2$ or anisotropy-parameter profiles decreasing with radius yield potential profiles differing strongly from the profile obtained from gravitational lensing. Best agreement with gravitational lensing is achieved with an anisotropy parameter increasing to $\beta\approx0.6$ within $0.5\,\mathrm{Mpc}$ and staying constant further out.
\end{itemize}

Our results show that the method returns convincing results on one well-studied galaxy cluster which we use here as a test case. Even though the uncertainties are still large, the anisotropy parameter can be rather well constrained by comparing our results with those obtained from gravitational lensing. We see three main future applications of this method: (1) in joint reconstructions of galaxy-cluster potentials compatible with all cluster observables; (2) in constraints of the anisotropy parameter; and (3) in tests of fundamental assumptions such as hydrostatic or virial equilibrium, and possibly also of general relativity, if applied to large and well-measured cluster samples.

\section{Discussion of other methods}
On top of the conditions leading to the Jeans equation, the method presented above implicitly rests on the assumption that all test particles (i.e. the galaxies) have the same mass, that their number density is proportional to their mass density and that one can establish a polytropic relation between their mass density $\rho$ and the effective galaxy pressure $P := \rho_\mathrm{gal}\sigma_r^2$. Fixing the velocity anisotropy profile $\beta(r)$ enables us to constrain the gravitational potential of the cluster via the radial Jeans equation using the observed density-weighted galaxy velocity dispersion. Our method does \emph{not} assume the mass to follow light a priori, i.e. $\rho_\mathrm{DM}$ does not necessarily need to follow $\rho_\mathrm{gal}$. Since the resulting potential is non-parametric, one can test the validity of particular parametric models as done above.

Another common deprojection method is the Abel inversion using derivatives of observables. However, due to the fluctuating nature of most observables which cause strongly fluctuating derivatives, we decided to use the Richardson-Lucy deprojection instead which only involves integrals.

There are many other methods to determine the gravitational potential of a cluster, or its mass, or to constrain the velocity anisotropy. For example, \cite{BinneyMamon1982} use the line-of-sight velocity dispersions as observables together with the surface brightness in spherical galaxies. Assuming a constant mass-to-light ratio, they can replace the unknown density in the Jeans equation by the luminosity, and can thus determine the velocity anisotropy profile $\beta(r)$. However, in the context of galaxy clusters it is not clear that the same relation between mass and luminosity should hold as well.

\cite{DejongheMerrit1992} determine the potential $\Phi$ by taking into account higher than second-order velocity moments of the collision-less Boltzmann equation (CBE). Given the velocity dispersions, and making use of a linear relation between the observables and $\Phi$, they expand the gravitational potential in terms of basis functions with unspecified coefficients. Using the positivity constraint of the distribution function and higher order moments of the CBE, they are able to determine the coefficients and thus the potential. However, from a practical point of view, it becomes increasingly difficult to constrain higher-order moments of the velocity distribution from their observable line-of-sight projections, as the projection integrals turn out to be substantially more complex with increasing order (see \cite{RichardsonFairbairn2014} for instance).

A method very similar to ours is proposed by \cite{MamonBoue2010}: starting again from the Jeans equation and fixing $\beta(r)$, they determine both the number density and the velocity dispersion observationally. Besides their using the Abel inversion, the only difference to our method is that they determine the number density of galaxies appearing in the Jeans equation by counting. Hence, they do not need our assumed polytropic relation between density and effective pressure. In future work, one could use this to test directly to what degree the polytropic assumption is appropriate.

\cite{Wolf2010} and \cite{vanderMarel2000} are considering a general parametrisation of $\beta(r)$ and of the mass $M(r)$ and constrain all free parameters via a maximum-likelihood analysis of the velocity dispersion predicted by the Jeans equation compared to the observed one. \cite{vanderMarel2000} use a constant velocity anisotropy whereas \cite{Wolf2010} take a varying parametric profile.

Another way to break the anisotropy mass degeneracy is taking the fourth order velocity moments of the CBE into account (\cite{Lokas2002}, see also \cite{RichardsonFairbairn2014} in this context). By expressing the velocity distribution function in terms of two integrals of motion, the energy $E$ and the angular momentum $L$, and further assuming that the distribution is separable in $E$ and $L$ with $\beta = \mathrm{const}.$, the two fourth-order moments are reduced to one equation involving the radial velocity to the fourth power, $\beta$, the radial velocity dispersion and $\Phi$. Thus, by measuring the velocity dispersion and $v_r^4$ one can infer the gravitational potential by solving this fourth-order equation together with the Jeans equation.

The MAMPOSSt method introduced by \cite{Mamon2013} breaks the degeneracy assuming parametric forms for the gravitational potential, the velocity anisotropy and the distribution of 3D-velocities. By performing a maximum-likelihood fit of the galaxy distribution in the projected phase space, all parameters can be determined. However, it has recently been shown (\cite{RichardsonFairbairn2014}) that their assumption of a Gaussian 3D-velocity distribution is incompatible with the equilibrium assumption underlying the CBE. One would thus have to choose a different 3D-velocity distribution.

\begin{acknowledgements}
This work was supported in part by the Collaborative Research Center TR~33 ``The Dark Universe'' of the Deutsche Forschungsgemeinschaft. This work is based on data collected at the ESO VLT (prog. ID 186.A-0798).
\end{acknowledgements}

\bibliographystyle{aa}
\bibliography{references}

\end{document}